\documentclass[fleqn,usenatbib]{mnras}

\usepackage{newtxtext,newtxmath}
\DeclareMathAlphabet{\mathpzc}{OT1}{pzc}{m}{it}
\usepackage[T1]{fontenc}

\DeclareRobustCommand{\VAN}[3]{#2}
\let\VANthebibliography\thebibliography
\def\thebibliography{\DeclareRobustCommand{\VAN}[3]{##3}\VANthebibliography}

\usepackage{graphicx}	
\usepackage{amsmath}	
\usepackage{amssymb}	

\title[Interstellar collisional desorption]{Icy molecule desorption in interstellar grain collisions}

\author[Kalv\=ans]{
Juris Kalv\=ans,$^{1}$\thanks{E-mail: juris.kalvans@venta.lv}
Kedron Silsbee,$^{2}$
\\
$^{1}$Engineering Research Institute “Ventspils International Radio Astronomy Center” (VIRAC)\\
of Ventspils University of Applied Sciences, In$\rm \check{z}$enieru 101, Ventspils, LV-3601, Latvia\\
$^{2}$ Max-Planck Institute for Extraterrestrial Physics, Giessenbachstrasse 1,
85748 Garching, Germany\\
}

\date{Accepted June 20, 2022. Received June 1, 2022; in original form November 13, 2021}

\pubyear{2021}

\begin{document}
\label{firstpage}
\pagerange{\pageref{firstpage}--\pageref{lastpage}}
\maketitle

\begin{abstract}
Observations of gaseous complex organic molecules (COMs) in cold starless and prestellar cloud cores require efficient desorption of the COMs and their parent species from icy mantles on interstellar grains. With a simple astrochemical model, we investigate if mechanical removal of ice fragments in oblique collisions between grains in two size bins (0.01 and 0.1 micron) can substantially affect COM abundances. Two grain collision velocities were considered -- 10 and 50 meters per second, corresponding to realistic grain relative speeds arising from ambipolar diffusion and turbulence, respectively. From the smaller grains, the collisions are assumed to remove a spherical cap with height equal to 1/3 and 1 ice mantle thickness, respectively. We find that the turbulence-induced desorption can elevate the gas-phase abundances of COMs by several orders of magnitude, reproducing observed COM abundances within an order of magnitude. Importantly, the high gaseous COM abundances are attained for long time-scales of up to 1 Myr and for a rather low methanol ice abundance, common for starless cores. The simple model, considering only two grain size bins and several assumptions, demonstrates a concept that must be tested with a more sophisticated approach.
\end{abstract}




\section{Introduction}
\label{intrd}

Desorption of molecules from interstellar grains is a key process that maintains molecules and atoms heavier than H or He in the gas phase even in long-lived dense interstellar cloud cores \citep{Leger83,Whittet10}. Besides thermal sublimation, desorption mechanisms thought to be operating in such cores include photodesorption, desorption by cosmic ray hits to interstellar grains, cosmic-ray-induced photodesorption, and reactive desorption.

Complex organic molecules (COMs) observed in cold cores \citep[e.g.][]{Bacmann12,Cernicharo12,Soma18} or their parent species (such as methanol) are known to be formed in the icy mantles adsorbed on to interstellar grains. However, astrochemical models considering the above-mentioned desorption mechanisms can hardly reproduce the observed abundances of COMs. There are hypotheses for such an effective non-thermal desorption mechanism \citep[e.g.][]{Vasyunin17,Harju20}. However, a credible and general answer to this problem remains elusive as of yet. With the help of an astrochemical model, we aim to contribute towards the understanding of non-thermal desorption in prestellar and starless cores by investigating the potential chemical effects of desorption induced by grain collisions.

Grain collisions in starless or star-forming cores can result in coagulation or shattering \citep{Scalo77,Kesselman78,Weidenschilling94}. The relatively low kinetic energies of grains in long-lived cores, along with observational evidence of grain growth implies that coagulation is the more relevant process in such cores \citep{Hirashita13}. Nevertheless, grain growth is accompanied with fragmentation \citep{Ormel09}. Regardless of their eventual outcome, grain collisions with velocities similar to or greater than a few m\,s$^{-1}$ have energies that exceed surface molecule desorption energies $E_D$ by orders of magnitude. This aspect means that grain collisions result in at least some desorption of icy molecules. 

For the aim stated above, we are interested in the frequency of grain collisions and the number of molecules removed from the icy mantle of a colliding grain. Other tasks for this study include developing a suitable astrochemical numerical model, as well as choosing a credible yet non-sophisticated approach for collisional desorption and appropriate values for parameters that govern this process.

\section{Methods}
\label{m_mtd}

The astrochemical model was designed to test the effects of collisional desorption in a simple and traceable manner. Macrophysically, we consider a stable, long-lived dense cloud core. To consider the chemical effects from collision between grains of different sizes, the simplest possible model with only two grain sizes (0.01 and 0.1\,$\mu$m) was employed (Section~\ref{m_phys}). In addition to gas-phase chemistry, surface and bulk-ice reactions were considered on both types of grains. Surface molecules or their fragments (chemical radicals) can be ejected into the gas phase by a number of photo- and chemical desorption mechanisms (Section~\ref{m_chem}). The small and the big grains can collide, inducing desorption of icy molecules from each other (Section~\ref{m_colli}). For the latter process, three cases were considered: `No-collision' (for reference), `Standard', and `Maximum' models. The Standard model investigates grain collisions induced by ambipolar diffusion at velocities of 10\,m\,s$^{-1}$, while the Maximum model considers turbulence-driven collisions at 50\,m\,s$^{-1}$ (Section~\ref{m_vrel}). For the small grains, these collisions result in the ejection of an ice fragment with volume equal to that of a spherical cap. For the Standard model, the height $h$ of the cap is equal to approximately 1/3 of the thickness of the icy mantle, while for the Maximum model $h$ equals ice thickness. The fragments were assumed to undergo subsequent destruction at a relatively short time-scale. For the big grains, it was assumed that their surface molecules are `scraped off' the grain individually, with desorption yield depending on molecular desorption energy (Section~\ref{m_cdes}).

\subsection{Physical model}
\label{m_phys}

For evaluating the efficiency of collisional desorption, the astrochemical code \textsc{Alchemic-Venta} was used. An exhaustive description of this model has recently been published by \citet{K21}. Here we provide only a basic description and detail the changes made for this study. As for an idea-testing study, the model was simplified to make the effects of collisional desorption easily discernible and reproducible.

A starless core with constant physical parameters was modelled. The relatively complex multilayer description of the icy mantles was left intact because it was useful for calculating collisional desorption yields. Grain collisions were described as encounters between grains of different sizes, which required adding at least a second one grain size bin to the model.

The \citet{K21} model was simplified by replacing the physical evolution of a star-forming cloud with with a pseudo time-dependent simulation of chemistry in unchanging physical conditions relevant for stable starless cloud cores. In addition to reproducibility, such a simple pseudo-time dependent model allows a simpler interpretation of results and often has served as a test-bed for investigations of astrochemical processes.

In order to consider collisions between grains of different sizes, grains with at least two sizes have to be included. We follow the practice established by \citet{Acharyya11}, where the bins are logarithmically spaced and the assumed average grain size has the integrated abundance of all grains in that bin. For reproducibility, we adopt the simplest possible approach, considering only large grains with radius $a=0.1\,\mu$m  and small $0.01\,\mu$m  grains, divided in bins with the MRN grain size distribution \citep{Mathis77}. This approach results in the 0.1\,$\mu$m grains having an abundance of $n_{\rm large}=1.66\times10^{-12}n_{\rm H}$ and for the small grains $n_{\rm small}=1.66\times10^{-10}n_{\rm H}$, where $n_{\rm H}$ is the number density of H atoms.

While more grain sizes certainly allow a more accurate modelling of gas-grain interactions and surface chemistry \citep{Pauly16,Iqbal18,Zhao18,Sipila20}, considering even only two grain sizes, large and small, is an improvement. For reproducibility, it is important that using only two grain sizes limits the number of grain collision types and the number of parameters to be estimated for calculating the collisional cross sections (Section~\ref{m_colli}).

The \textsc{Alchemic-Venta} model considers grain growth due to ice mantle accumulation. Less than 30 icy molecule monolayers (MLs) or up to $\approx0.01$\,$\mu$m thick ice layer accumulate on to both grain types. While this is a relatively small increase for the 0.1\,$\mu$m grains, it significantly increases the cross section and surface area available for for the 0.01\,$\mu$m grains, which has an effect on the gas-grain chemistry \citep{Acharyya11}.

The growth or shattering of grains due to collisions was not considered in order not to mix different effects and allow a clear interpretation of collisional desorption effects. Moreover, the consideration of these processes would add poorly-known variables (e.g., sticking and shattering cross sections for icy grains), which can complicate the results. However, taking into account that grains generally tend to coagulate in dense cores, we consider an integration time of only up to 1\,Myr, at which the collision-induced changes in grain size distribution can become significant \citep{Ormel09,Silsbee20}.

The model considers a plane-parallel molecular cloud (slab) illuminated by the interstellar radiation field from both sides with a visual extinction $A_V=10$\,mag. Cloud density was taken to be $n_{\rm H}=2\times10^4$\,cm$^{-3}$. Grains of different sizes have different temperatures, which affects surface chemistry and the resulting composition of ices \citep{Pauly16,Iqbal18,Sipila20}. For simplicity, here a uniform temperature of $T=10$\,K was assumed for the large grains, small grains, as well as the gas. The standard cosmic ray total ionization rate $\zeta=1.36\times10^{-17}$ of the UDfA12 network was employed.

\subsection{Chemical model}
\label{m_chem}

\begin{table}
\begin{center}
\caption{Initial chemical abundances relative to hydrogen.}
\label{tab-ab}
\begin{tabular}{lc}
\hline\hline\\
Species & Abundance \\
\hline
H$_2$ & 0.500 \\
He & 0.0900 \\
C$^+$ & $7.30\times10^{-5}$ \\
N & $2.14\times10^{-5}$ \\
O & $1.76\times10^{-4}$ \\
F & $6.68\times10^{-9}$ \\
Na$^+$ & $2.25\times10^{-9}$ \\
Mg$^+$ & $1.09\times10^{-8}$ \\
Si$^+$ & $9.74\times10^{-9}$ \\
P$^+$ & $2.16\times10^{-10}$ \\
S$^+$ & $9.14\times10^{-8}$ \\
Cl & $1.00\times10^{-9}$ \\
Fe$^+$ & $2.74\times10^{-9}$ \\
\hline
\end{tabular} \\
\end{center}
\end{table}

The chemical model was adopted from \citet{K21}, again with several simplifications for result clarity and reproducibility. The reaction list is based on the UDfA12 database \citep{McElroy13} for gas phase and the OSU network \citep{Garrod08} for surface chemistry. Surface molecule $E_D$ were updated with values from \citet{Wakelam17}. The initial chemical abundances are listed in Table~\ref{tab-ab}. Similarly to \citet{Acharyya11}, each grain size has its own surface reaction network.

With the multilayer approach, four layers of icy mantles were modelled -- surface and three subsurface bulk-ice layers \citep[`sublayers'; see][for a detailed description]{K15apj1}. In addition to collisions, several desorption mechanisms were included -- thermal sublimation, photodesorption by interstellar photons, cosmic-ray-induced desorption, cosmic-ray-induced photodesorption, reactive desorption, desorption of surface molecule photodissociation products, and indirect reactive desorption. The latter was included again with the approach of Model~F of \citet{K15apj1} in the \textsc{Alchemic-Venta} model thanks to the results by \citet{Pantaleone21}, who showed that the energy released by the exothermic H+H surface reaction may result in the desorption of nearby icy molecules. The Model~F approach assumes uniform desorption yield for all icy species with desorption energy $E_D$ up to 2600\,K, helping to produce an interstellar ice composition that is consistent with observations of dense cores, such as \citet{Oberg11}.

The detailed approach on cosmic-ray-induced desorption of the \citet{K21} model, involving a separate phase of heated icy grains, was dropped. Instead, the rate of this process was calculated with the simpler and trusted approach of \citet{Hasegawa93}, which describes molecules sublimating from an icy grain suddenly heated to 70\,K by a cosmic ray impact. The rate coefficient is proportional to the frequency of CR hits heating grains to 70\,K $f_{70}=3.16\times10^{-14}$\,s$^{-1}$ \citep[taken directly from][]{Hasegawa93} and the cooling time $t_{\rm cool}$. In \citet{K21}, the latter was calculated with an accurate method, which is no longer valid here, because of the simplified model. Therefore, we take $t_{\rm cool}$ equal to the characteristic time-scale for icy CO sublimation ($1.3\times10^{-4}$\,s in this model) as proposed by \citet{Hasegawa93}. The cooling time of the two grain types is similar because $t_{\rm cool}$ depends on the number of monolayers from which the volatile CO molecules are removed, not the number of molecules sublimated or available on the surface \citep{KK20ii,K21}.

A problem is to calculate the cosmic-ray-induced heating rate of the small grains in a simple yet reliable way. This problem is made complex by the icy mantle, which may constitute most of the mass and volume of the 0.01\,$\mu$m grains. Generally, cosmic-ray impacts can heat small grains to higher temperatures more often than large grains. However, this effect is reduced by the accumulation of the icy mantle, which has a similar heat capacity to the refractory grain nuclei but is less dense and thus absorbs less energy from the cosmic-ray particles. Here, to find $f_{70}$ for the small grains, we extrapolate the data of \citet{K16}. This work showed that the grain heating rate to or above a temperature threshold is approximately equal for grains, large and small, having a similar ice mantle thickness. Therefore, for the small grains, we adopt the same $f_{70}$ as for the large grains.

\subsection{Grain collisions}
\label{m_colli}

\begin{figure}
		\vspace{3cm}
	\includegraphics[width=10cm]{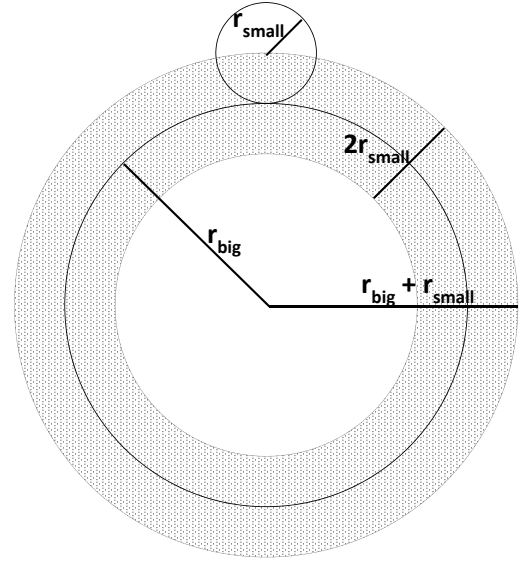}
		\vspace{-4cm}
		\caption{Schematic representation of Equation~(\ref{colli4}): the cross-section (shaded area) for grain collisions inducing desorption. The total radius of small and big grains, is abbreviated as $r_{\rm small}$ and $r_{\rm big}$, respectively.}
    \label{fig-S}
\end{figure}
In this simple study, we test the chemical effects of grain collisions with the relative velocity $v$ between differently-sized grains as the main changing parameter. The relative velocity regulates collision frequency and energy. The latter determines the desorption yield. A straightforward way to estimate the yield is to assume that some of the collisional energy is transformed into heat, which subsequently induces sublimation of ices, as mentioned by \citet{Sipila20}. However, thermal desorption has little means for removing methanol or other icy COMs observed in dense cores because the heated grain rapidly cools down by sublimating only the most volatile molecules, such as CO. Even without volatiles, excessively high temperatures (several hundreds of K) are required to appreciably sublimate molecules with desorption energies comparable to that of water \citep{KK20ii}.

Taking the above into account, we propose and apply a different desorption mechanism -- mechanical removal of ice layers in oblique collisions. Such events may result in grains slipping and rolling off each other, partially retaining their momentum, which means that each grain receives less energy from the collision. This process is determined by collision velocity, angle, and surface properties of both grains. Moreover, the latter properties change with time, starting with bare grains at the start of the interstellar cloud simulation, continuing with water ice covered grains and probably ending with a CO-dominated ice surface layer. Therefore, oblique collisions are complex events with a variety of possible outcomes. Here, we are interested only in a single aspect of the oblique collisions -- pushing off the relatively weakly-bound icy mantles from the grain. Such removal may be less possible in head-on collisions, hence the importance of colliding obliquely.

Contrary to sublimation, this mechanism does not require that each intermolecular bond is split during the collision. Instead, molecules can be removed in the form of tiny fragments made up entirely of ice. The fragments then disintegrate due to interaction with photons and cosmic-ray protons. Here we assume that such two-step desorption does not affect the overall desorption yield; it is discussed further in Section~\ref{m_frag}.

The collisional desorption rate coefficient for icy species can be expressed as
\begin{equation}
    k_{\rm coll} = RY,
	\label{colli1}
\end{equation}
where $R$ is the collision rate and $Y$ is the desorption yield, discussed in Section~\ref{m_cdes}. The collision rate (or frequency) per grain characterises how often a single grain experiences a collision with a grain of a different size that results in removal of icy matter. The collision rate for a large grain (colliding with the small grains) is
\begin{equation}
    R_{\rm large} = S v n_{\rm small},
	\label{colli2}
\end{equation}
where $S$ is the collision cross section, $v$ is the relative velocity between the colliding grains, and $n_{\rm small}$ is the numerical abundance of the small grains that impact of the large grain. Complementary, the small grains experience collisions with a frequency 
\begin{equation}
    R_{\rm small} = S v n_{\rm large},
	\label{colli3}
\end{equation}
where $n_{\rm large}$ is the numerical abundance of the large grains. Equations (\ref{colli2}) and (\ref{colli3}) mean that the collisional desorption rate of icy molecules residing on a certain grain depends on the abundance of impactor grains.

The grain relative velocity was calculated in Section~\ref{m_vrel}. For the collision cross-section, Figure~\ref{fig-S} schematically shows that an oblique collision occurs when the small grain hits an edge of the large grain that has its width  equivalent to the small grain radius:
\begin{equation}
    S = \pi(a_{\rm l}+b_{\rm l}+a_{\rm s}+b_{\rm s})^2-
		\pi(a_{\rm l}+b_{\rm l}-a_{\rm s}-b_{\rm s})^2\,.
	\label{colli4}
\end{equation}
Parameters $a_{\rm l}, b_{\rm l}, a_{\rm s}$, and $b_{\rm s}$ are the radius and ice thickness of the large grains and the radius and ice thickness of the small grains, respectively. This estimate of $S$ is based on logical considerations, just as the cross-section assessment for grain coagulation \citep{Ormel09}.

\subsection{Collision velocities}
\label{m_vrel}

The prominent infall motions in prestellar cores have velocities of $10^2...10^3$\,m\,s$^{-1}$ \citep{Tafalla98,Caselli02,Schnee13,Chung21}. For quiescent starless cores, the velocities of gas flows due to several mechanisms are closer to the $10^2$\,m\,s$^{-1}$ \citep{Pavlyuchenkov06,Pineda10,Ohashi16}. Here we consider the case of a presumed typical quiescent core.

As noted before, to explore the significance of collisional desorption, three separate cases were considered -- `No-collision', `Standard', and `Maximum' models. The No-collision case serves as a reference model. The Standard model is intended to consider grain collisions with velocities $v$ that are likely to occur in common quiescent starless cores. The Maximum model considers collisions with relative velocities that are realistic yet close to the maximum values available for starless and prestellar cloud cores. In the Maximum model, the core is assumed to be affected by an external influence that induces mildly supersonic gas motions.

To obtain $v$ for the Standard model, we employ the approach of \citet{Silsbee20} taking a lower density and somewhat larger core than L1544, with $n_c=2\times10^4$\,cm$^{-3}$ and $r_0=5000$\,AU (see Equation~(1) in that paper). We take an ionization fraction of $6\times10^{-8}$ (as calculated with our chemical model at $t=1$\,Myr) and magnetic field strength of 10\,$\mu$G \citep{Crutcher10}. Equation~(5) of \citet{Silsbee20} was rewritten as
\begin{equation}
    c_d\mathpzc{g}\rho_\mathpzc{g} = \sum_i{n_iv_{\rm in}[\mu^i_{\rm H_2}n_{\rm H_2}\left\langle \sigma v \right\rangle_{\rm H_2} + \mu^i_{\rm He}n_{\rm He}\left\langle \sigma v\right\rangle_{\rm He}]},
	\label{colli5}
\end{equation}
where $\mathpzc{g}$ is the local gravitational field; $\mu_{\rm H_2}$ and $\mu_{\rm He}$ are the reduced masses of molecular hydrogen and helium, respectively, in collisions with the ions; and $\left\langle \sigma v \right\rangle$ is the ion-neutral momentum transfer cross section averaged over a Maxwellian velocity distribution'' \citep[][p.~A39]{Silsbee20}. Equation~(\ref{colli5}) was calculated for a sum over the ions in the gas, as calculated in the model. The most abundant ions are H$_3^+$, He$^+$, C$^+$, and H$^+$.

With these parameters, we get a grain drift rate of 15\,m\,s$^{-1}$, which translates in collision velocities arising from ambipolar diffusion of 15\,m\,s$^{-1}$ for 0.01 and 0.1\,$\mu$m grains. This assumes that the parameter $c_d$ in Equation~(5) of \citet{Silsbee20} is taken to be unity. This collision speed is only equal to the drift rate because we took one grain in the size range which is magnetically coupled, and one which is more strongly coupled to the neutral gas. If instead we had taken 0.001 and 0.01\,$\mu$m grains, the collision speed would only be 1\,m\,s$^{-1}$, or if we had taken 0.1 and 1.0\,$\mu$m grains, it would be just 3\,m\,s$^{-1}$. Considering 15\,m\,s$^{-1}$ as an upper limit, we take $v=10$\,m\,s$^{-1}$ for the Standard model.

For the Maximum model, $v$ was calculated with the other potentially significant source of grain relative velocity -- the motions of particles in response to turbulence. The strength of the turbulence present is of course a tricky question. Assuming the turbulence to have a sonic Mach number $\mathcal{M}$ at the scale of $r_0$, for a temperature of 10\,K, we find that the collision speeds are $20\mathcal{M}^{3/2}$\,m\,s$^{-1}$.

We begin with Equation~(28) from \citet{Ormel07}. This shows that assuming the Stokes numbers for both grains are small, the collision speed is given by
\begin{equation}
v_{\rm coll} = \chi(\epsilon)\sqrt{St_L} V_g 
\label{eq:OAndC28}
\end{equation}
where $St_L$ is the stokes number of the larger grain, $V_g$ the turbulent velocity, and $\chi$ is a number between 1.4 and 1.7 which depends on the ratio of Stokes numbers between the large and small grain.  The Stokes number is defined as the stopping time $t_s$ of the large grain divided by the eddy turnover time $t_L$ at the largest scale $r_0$.

These grains are in the Epstein drag regime where the stopping time is given by
\begin{equation}
t_s = \frac{3m}{4v_{\rm th} \rho_g \sigma_g}
\end{equation}
where $m$ is grain mass, $v_{\rm th} = 0.92 \sqrt{4k_B T/(\pi m_p)}$ is the thermal velocity scale of the gas (with a factor to account for the helium fraction, $\rho_g$ is the gas density, and $\sigma_g =\pi a^2$ the collision cross-section.  This assumes that all of the hydrogen is molecular, and the medium contains 1 helium atom per 10 hydrogen atoms.

The eddy turnover time at the largest scale $r_0$ can be calculated from Equation~(2) of \citet{Ormel07}, the assumption that the turbulent energy spectrum $E(k) \propto k^{-5/3}$ and the relation $t_k = r_0/V(k)$, with $V(k) = \sqrt{2kE(k)}$. This gives
\begin{equation}
t_L = \sqrt{3/2}\,r_0/V_g.
\end{equation}
Putting this together, we find that the Stokes number of a grain with size $a$ and average density $\rho_d$ is
\begin{equation}
St = 0.52 \frac{\rho_d a}{\rho_g r_0} \frac{V_g}{c_s}.
\end{equation}
where $c_s = \sqrt{k_B T/(2.33 m_p)}$ is the (isothermal) sound speed in the gas, and from Equation~(\ref{eq:OAndC28}),
\begin{equation}
v = \chi(\epsilon) \sqrt{\frac{.52\rho_d a c_s^2}{\rho_g r_0}} \mathcal{M}^{3/2} \equiv \psi \mathcal{M}^{3/2},
\end{equation}
where $\mathcal{M} = V_g/c_s$ is the isothermal Mach number for turbulence at the scale of $r_0$.

For a temperature of 10\,K, we find that the collision speeds are $20\mathcal{M}^{3/2}$\,m\,s$^{-1}$. For example, for $\mathcal{M}=2$, the collision speed of 0.01 and 0.1\,$\mu$m grains $v = 60$\,m\,s$^{-1}$. For the Maximum model, a core with a slightly supersonic turbulence was considered with $v=50$\,m\,s$^{-1}$.

\subsection{Collisional desorption yield}
\label{m_cdes}

\begin{figure}
		\vspace{2cm}
	\includegraphics[width=10cm]{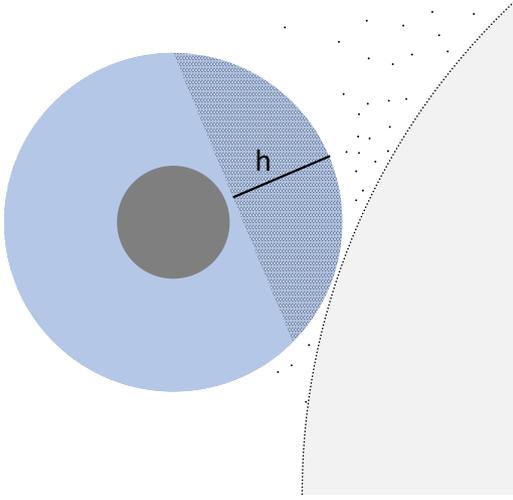}
		\vspace{-3cm}
		\caption{Schematic drawing of ice removal in a collision between small and large grains. From the small grains, an icy spherical cap (shaded) with height $h$ is removed, while separate surface molecules are detached from the large grain.}
    \label{fig-h}
\end{figure}

The desorption yields (i.e., volume of detached ice fragments) differ for the Standard and Maximum models, owing to their different collisional energies. As stated in Section~\ref{m_phys}, we do not follow the fate of the collided grains, e.g., their sticking or heating, focusing only to the mechanical desorption effects.

In order to estimate desorption yields, we consider dust grains at freeze-out conditions, when both the 0.1 and 0.01\,$\mu$m grains have $\approx0.01\,\mu$m thick icy mantle. The fragmenting of such a thick mantle requires higher energies than for a thin mantle. The density of dust nuclei was taken to be 3\,g\,cm$^{-3}$, while that of ice -- 1\,g\,cm$^{-3}$. The collision energy available for splitting off the ice fragment is
\begin{equation}
  E_{\rm avail} = 0.5\mu v^2 X_{\rm des},
	\label{cdes1}
\end{equation}
where $\mu$ is the reduced mass and $X_{\rm des}$ is efficiency parameter. With little knowledge on the detailed physics of oblique collisions of differently-sized icy grains, here we simply assume that the proportion of collisional energy available for detaching ice mantle fragments $X_{\rm des}\leq0.1$. Clarification of this proportion needs microphysical studies and parameter space investigation, neither of which is within the scope of this study. We assume that capping the energy yield at 10 per cent is a sufficiently cautious approach. With $X_{\rm des}=0.1$ we mean that $\leq10$ per cent of collision energy are consumed in desorption from the large grains and $\leq10$ per cent from the small grains, i.e., the total proportion of energy used in desorption is up to 20 per cent. However, practically $X_{\rm des}$ can also be assumed to be $\leq10$ per cent because the contribution of desorption from the large grains is minor.

\subsubsection{Desorption from large grains}
\label{m_Ylarge}

The collisions do not have sufficient energy to remove a multi-monolayer (ML) fragment from the large grains. Therefore, we assume that the small grain is only scraping the surface of the large grain, which results in the detachment of individual surface molecules. At the maximal 50\,m\,s$^{-1}$ velocity, $E_{\rm avail}=33$\,eV, which corresponds to the desorption of about 300 CO molecules (assuming desorption energy $E_{D,\rm CO} = 1300$\,K). This is a tiny faction of the surface area of a 0.1\,$\mu$m grain with an icy mantle. The yield of such desorption is
\begin{equation}
  Y_{\rm large} = \frac{E_{\rm avail}}{\bar{E}_D},
	\label{cdes2}
\end{equation}
where $\bar{E}_D$ is an assumed average desorption energy of the outer surface molecules. Its is possible to calculate $\bar{E}_D$ precisely and correctly by averaging $E_D$ of surface-layer molecules. However, this composition can vary significantly even for similar models. Therefore, again for traceability and reproducibility of the model, we simply assume that $\bar{E}_D = 2E_{D,\rm CO}$. Equation~(\ref{cdes1}) is applied in both, Standard and Maximum models. For practical purposes, in the calculations code $Y_{\rm large}$ was expressed as a fraction of the total number of molecules in the surface layer. Summarizing, oblique small grain impacts on large grains results in unselective removal of a limited number of molecules from the surface layer, while bulk ice remains unaffected. Because $Y_{\rm large}$ is so small, desorption from the large grains is negligible and has little effect on the results. The exact yields for different molecules depend on the composition of the surface layer of the big grains, which is constantly evolving. A characteristic composition the Standard model with a fully-formed $\approx25$\,ML ice mantle is CO and H$_2$O (40 per cent each), along with a few per cent of other species, such as NH$_3$, O$_2$, CH$_4$, H$_2$O$_2$.

\subsubsection{Desorption from small grains}
\label{m_Ysmall}

For the small grains, we assume that the volume of an ice fragment that can be split off is equal to a spherical cap with a height $h$, as shown in Figure~\ref{fig-h}. This assumption makes sense of considering the oblique collisions, where a fragment roughly in the shape of a spherical cap can be shovelled or shifted off the grain. This mechanism also suggests that any excess energy available for desorption in the cases of high $v$ or high $X_{\rm des}$ will have a little effect on the yield $Y_{\rm small}$ and thus the approach is somewhat resilient to changes in parameters. The volume of the the cap is
\begin{equation}
  V = \frac{\pi h^2 (3(a_s+b_s)-h)}{3},
	\label{Ncdes2}
\end{equation}
where $h$ is the height of the cap and $a_s+b_s$ is the total radius of the small grain. The number of molecules in the cap is equal to $V/a_m^3$, where $a_m=0.32$\,nm is the assumed size of a `cubic average' molecule.  

Assuming $h\geq 1/3 b_s$, 4 per cent or more of the ice mantle can be broken off. This amounts for an ice chunk consisting of at least $1.6\times10^4$ molecules. This process can be described in macrophysical terms, employing the tensile strength $\sigma_t$ of ice. Its value has been deduced to be on the order of 1\,MPa \citep{Petrovic03}. This value is slightly higher than that from the results from interaction between icy grain agglomerates \citep{Blum08}. The tensile strength increases with smaller length scales and the fractures occur between separate ice grains. Here we are interested in splitting small grains, which suggest a higher $\sigma_t$. On the other hand, interstellar ices have a large proportion of small non-polar molecules, as well as larger organic species. These and other irregularities will inevitably result in weakening the ice structure either by forming sub-grains that can be easily separated or by weakening the overall ice $\sigma_t$. Here we are unable to quantitatively assess these effects and we adopt $\sigma_t$ based on the published values.

For the Standard model, we assume that collisional desorption affects the outer third of the icy mantle, i.e., $h\approx0.003\,\mu$m. In the chemical model, this means that out of the four ice mantle layers, only the thin surface layer ($\approx1$\,ML) and the outer bulk-ice layer are affected. In the case of a near-complete freeze-out, these two layers constitute about one third of the ice mantle thickness (more in volume because outer MLs have more molecules). About 40 per cent of this part of mantle consists of the nonpolar CO, CO$_2$, and N$_2$ molecules (cf. Figure~\ref{fig-major}). Such a polar-nonpolar mixture can be expected to have weak structural properties. Therefore, it may be reasonable taking $\sigma_t=1.0$\,Mpa. The collision impact overcomes the tensile strength specifically in the plane of the flat cut area of the spherical cap of icy mantle. This area is equal to
\begin{equation}
  S_{\rm cap} = \pi [(a_s+b_s)cos\theta]^2\,,
	\label{cdes3}
\end{equation}
where $\theta$ is the angle between grain radius ending at the edge of the flat cut area, and the radius of the cut area, itself a circle.

Assuming that the shear involves two molecule monolayers, we obtain that the critical collision velocity for removal of a spherical cap with $h\approx0.003\,\mu$m and $X_{\rm des}$ is 7.4\,m\,s$^{-1}$. Therefore, such a removal is possible in the Standard model if we assume $\sigma_t$ similar to that given by \citet{Petrovic03}.

The Maximum model describes a case of high relative velocities between the large and small grains. Collisions occur more often and they are more violent. The higher collisional energy can be expected to allow splitting ice that is deeper in the mantle with a higher proportion of H$_2$O and higher tensile strength. For example, if we assume $\sigma_t=5$\,MPa (with two ice MLs involved in the shear), the critical velocity for the separation of the spherical cap with $h=0.01\,\mu$m is 38\,m\,s$^{-1}$. Therefore, it is reasonable to assume that a spherical cap with $h$ equal to full ice mantle depth occurs in the Maximum model.

The removal of ice fragments requires much less energy than complete sublimation of the icy molecules in the cap. With the assumptions for the Maximum model, splitting off the spherical cap consumes only 19\,eV, while instant vaporization of the same amount of ice requires $10^4...10^5$\,eV, depending of the $E_D$ of its molecules. Such a difference is what may make mechanical desorption efficient, despite consuming only a faction of collisional energy. Moreover, \textit{all} molecules, including COMs, in the spherical cap are removed, not only volatiles as in the case of grain heating.

Compared to other desorption mechanisms, collisional mechanical desorption is peculiar in that its efficiency grows substantially with each ice ML added to the small grain. The increase of the icy mantle increases the collision rate by increasing $S$ and the desorption yield by increasing $h$. As a result, for the Maximum model, the collisional desorption rate coefficient for a 0.01\,$\mu$m grain with 25\,MLs of ice at the end of the simulation is more than 100 times higher than that for the same grain with 1\,ML of ice at the start of the simulation. At the same, the total abundance of icy molecules has also grown by a factor of 50, giving an impressive $\approx5000$-fold increase in efficiency. Therefore, collisional desorption from the small grains is negligible at the start and becomes increasingly efficient until 0.25\,Myr, when the freeze-out process ceases and the mantles attain maximum thickness.

\subsection{Fate of removed ice fragments}
\label{m_frag}

We assume that the icy fragments ejected from grains in collisions disintegrate into separate molecules within a time-scale that is short compared to their ejection time-scale. The destruction of the fragments was not explicitly described in the code and the desorbed icy species were transferred directly to the gas phase. This is not a problem for desorption from the large grains, where the small impactor grains are assumed to scrape off individual surface molecules.

The disintegration of icy fragments removed from the small grains must be justified separately for the Standard and Maximum models. In the Standard model, the maximum characteristic size of the icy particle is about a third of the ice thickness near freeze-out conditions or $\approx3$\,nm. Such tiny particles can be heated significantly by a single UV photon, inducing sublimation of volatile molecules, such as CO \citep{Hoang20}. Sublimation reduces the size of the particle and may cause it to fragment, so that hits by subsequent photons heat the remnants to higher temperatures, sublimating H$_2$O and other less-volatile ices. We estimate that a cosmic-ray-induced photon will be able to hit such a particle within a few thousand years. This is orders of magnitude shorter than the time between grain collisions that can deplete the fragments before their molecules are sublimated.

The case is different in the Maximum model, where icy particles with a characteristic size of 10\,nm can be ejected. These particles largely consist of ices from inner ice MLs that contain less CO and more H$_2$O, CO$_2$, and other less-volatile species. UV photons are not able to heat significantly grains in this size range \citep{Cuppen06}. The 10\,nm icy particle may be hit by a cosmic-ray proton. At these length scales, the interaction of the fast proton with the grain is stochastic, and its prediction requires a complex model, such as that of \citet{Shingledecker17}. However, protons with energies sufficient to lift the grain temperature to hundreds of K are encountered with a frequency of about $10^{-11}$\,s$^{-1}$ \citep[as estimated with the methods of][]{K16}. While  cosmic-ray induced sublimation would cause further fragmentation of the particle, it is possible that the lifetime of the fragments is sufficiently long to affect the eventual desorption yield. Recognizing this lack of knowledge, a study on the stability and thermal sublimation of small ice grains is planned in the future with the \textsc{tcool} model of \citet{KK20ii}.

\section{Results}
\label{rslt}

In this section, we briefly discuss the general results of the model before exploring the effects of collisional desorption on the gas-phase abundances of COMs.

\subsection{General results}
\label{r_gen}

\begin{figure*}
	\includegraphics[width=20cm]{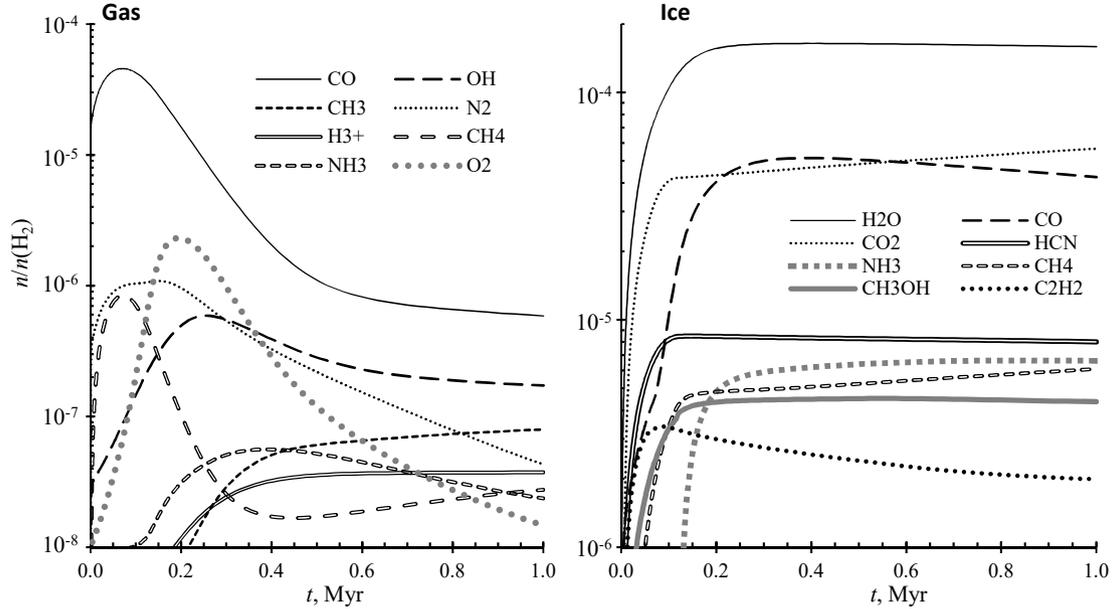}
		\vspace{-19cm}
		\caption{Calculated abundances relative to H$_2$ of major gas and solid ice phase molecules in the Standard model.}
    \label{fig-major}
\end{figure*}

A gas-phase chemical pseudo-equilibrum is achieved within an integration time of $t\approx0.1$\,Myr, at which point $\approx3$\,MLs of ice is already deposited on to the grains. Rapid freeze-out continues until $t\approx0.2$\,Myr. Both grain types attain ice thickness of about 25\,ML, while 66 per cent of the icy molecules are on the small grains, owing to their high abundance.

Figure~\ref{fig-major} shows the calculated abundances for major gas and icy species. These abundances do not differ significantly between the three simulations and, for clarity, only the results of the Standard model are shown. The primary gas-phase molecule is CO, followed by simple hydrocarbons, N$_2$, and O$_2$. A slow depletion of gaseous species continues for the whole duration of the simulations, never achieving a real adsorption-desorption equilibrium. Only 0.5 per cent of CO remains in the gas after 1\,Myr \citep[assuming that CO freezes out in the form of CO and CO$_2$,][]{Whittet10}. Given the model's simplicity, the calculated abundances of icy species are in a surprisingly good agreement with observations of starless cores \citep{Boogert11}, with the ratio H$_2$O:CO:CO$_2$:NH$_3$:CH$_4$:CH$_3$OH being 100:27:36:4:4:3 at $t=1$\,Myr.

\subsection{Results for complex organic molecules}
\label{r_com}

\begin{figure*}
	\includegraphics[width=20cm]{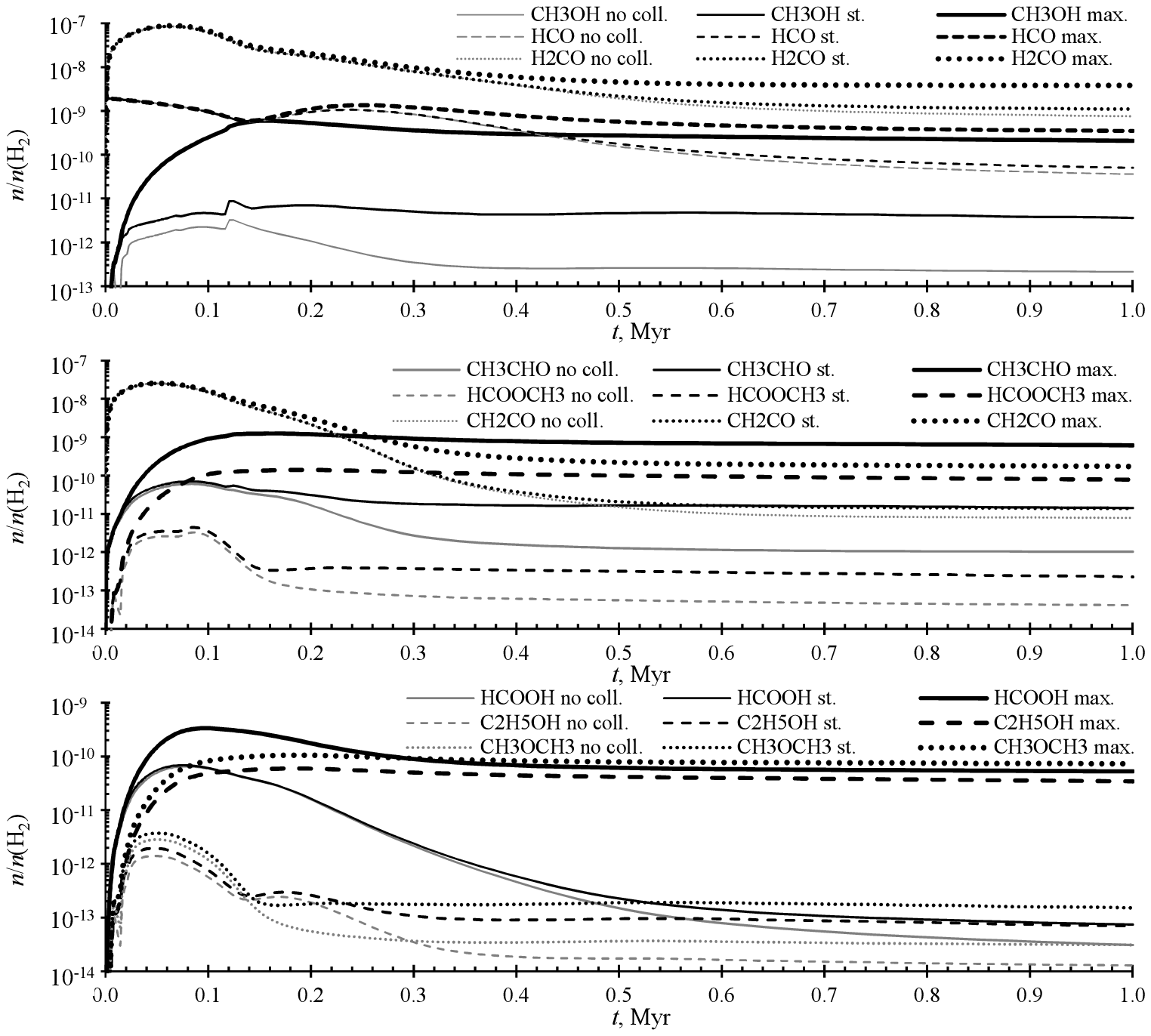}
		\vspace{-13.5cm}
		\caption{Calculated abundances relative to H$_2$ of COMs and other organic molecules in the No-collision, Standard, and Maximum models. We do not employ a logarithmic $y$-axis in order not to give the reader a visual impression that abundances calculated for the peak at $t\approx0.1$\,Myr are statistically comparable to abundances calculated for longer time-scales, when the relaxation of the model from the initial conditions has occurred.}
    \label{fig-com}
\end{figure*}

\begin{table*}
	\caption{Calculated relative abundances in the No-collision, Standard and observed COM gas-phase abundances.}
	\label{tab-com}
	\begin{tabular}{lcllllllll}
		\hline
 & \multicolumn{3}{c}{Abundances at 0.5\,Myr} & \multicolumn{3}{c}{Abundance ratio} &  &  \\
Species & No-collision & Standard & Maximum & St./No-coll. & Max./No-coll. & Max./St. & Observed abundances & References \\
		\hline
CH$_3$OH & 2.6(-13)$^a$ & 4.7(-12) & 2.7(-10) & 18 & 1000 & 59 & 1.2(-9)$^{b}$ & SS \\
H$_2$CO & 1.9(-9) & 2.1(-9) & 4.5(-9) & 1.1 & 2.4 & 2.1 & 4.0(-10); 1.3(-9) & M B \\
CH$_3$CHO & 1.3(-12) & 1.7(-11) & 7.1(-10) & 13 & 560 & 43 & 1.3(-10)$^{b}$ & SS \\
HCOOCH$_3$ & 5.5(-14) & 3.2(-13) & 9.9(-11) & 5.7 & 1800 & 310 & 2.0(-11); 7.4(-10); 1.5(-10) & C B J \\
CH$_3$OCH$_3$ & 3.6(-14) & 1.9(-13) & 7.9(-11) & 5.2 & 2200 & 420 & 2.0(-11); 1.3(-10); 5.1(-11) & C B J \\
CH$_2$CO & 1.4(-11) & 2.0(-11) & 2.2(-10) & 1.4 & 15 & 11 & 1.3(-11); 2.0(-10) & C B \\
CH$_3$O & 3.3(-14) & 2.3(-13) & 1.3(-11) & 7.0 & 380 & 54 & 4.7(-12); 3.3(-11); 2.7(-11) & C BF J \\
HCO & 1.5(-10) & 1.7(-10) & 5.6(-10) & 1.1 & 3.8 & 3.3 & 3.6(-10) & BF V \\
C$_2$H$_5$OH & 1.7(-14) & 9.6(-14) & 4.2(-11) & 5.5 & 2400 & 430 & ... & ... \\
		\hline
	\end{tabular}
\\
$^a$ $A(B)$ means $A\times10^B$.\\
$^b$ Median value, varies within a factor of 3.\\
$^c$ Median value, varies within a factor of 5.\\
References: SS: \citet{Scibelli20}; M: \citet{Marcelino05}; B: \citet{Bacmann12}; C \citet{Cernicharo12}; J: \citet{Jimenez16}; BF: \citet{Bacmann16}; V: \citet{Vasyunin17}.
\end{table*}

Figure~\ref{fig-com} shows the principal result of our study -- the abundances of COMs and associated simpler organic molecules, as calculated with the three simulations. Table~\ref{tab-com} shows that, compared to the No-collision model, the COM abundances increased by a factor of about 10 in the Standard model, while for the Maximum model they increased by 1-4 orders of magnitude. In absolute numbers, the most significant abundance increase is for formaldehyde, although H$_2$CO has an ice abundance that is about 50 times lower than that of methanol and even a few times lower than that of the more complex COMs, such as methyl formate (cf. Figure~\ref{fig-ice}). Nevertheless, the gas-phase abundance of H$_2$CO exceeds that of CH$_3$OH by more than an order of magnitude in the Maximum model. Observations of dense cores show that methanol and formaldehyde have comparable abundances (Table~\ref{tab-com}).

In the collisional desorption models, formaldehyde is overproduced at the expense of methanol, which has been desorbed directly from ices. This happens because of apparently over-effective CH$_3$OH gas-phase destruction in the UDfA12 network. In particular, CH$_3$OH reactions with H$^+$, CH, and other reactants disassemble the CH$_3$O structure, partially redistributing it to the aldehydes H$_2$CO and CH$_3$CHO. These reactions are not included in reaction networks based on the OSU database, such as \citet{Garrod08} or \citet{Vasyunin13}. The calculated increase of H$_2$CO abundance from the No-collision to Maximum models is $>10^{-9}$, similar to observed methanol abundances in starless cores \citep{Scibelli20}. From this value one can deduce that, apparently, collisional desorption in the Maximum model is able to provide a sufficient flux of methanol into gas to account for the production of COMs at a level that agrees with observations.

Despite the above discrepancy, in the Maximum model methanol has a final ($t=1$\,Myr) abundance of $2.1\times10^{-10}$ relative to H$_2$ and maximum abundance $6.0\times10^{-10}$ at 0.2\,Myr, an increase by a factor of $\approx10^3$. More complex COMs, such as HCOOCH$_3$ and CH$_3$OCH$_3$, are less affected by gas-phase destruction and have Maximum model gas-phase abundances very close to observed values. For these species, the increase of abundance is highest, amounting to around 2000 relative to the No-collision model. Regarding the contribution of collisional desorption to gas phase abundance, organic species fall in two distinct groups. For the simpler molecules HCO, H$_2$CO, and CH$_2$CO it is about a factor of few, while for the COMs consisting of 6 or more atoms and containing the methanol's CH$_3$O group, more than 99.5 per cent of gas-phase molecules arise from collisional desorption. Carbon chains largely remain unaffected, owing to their gas-phase origin.

\subsection{Ice layer -- the source of COMs}

\begin{figure*}
	\includegraphics[width=20cm]{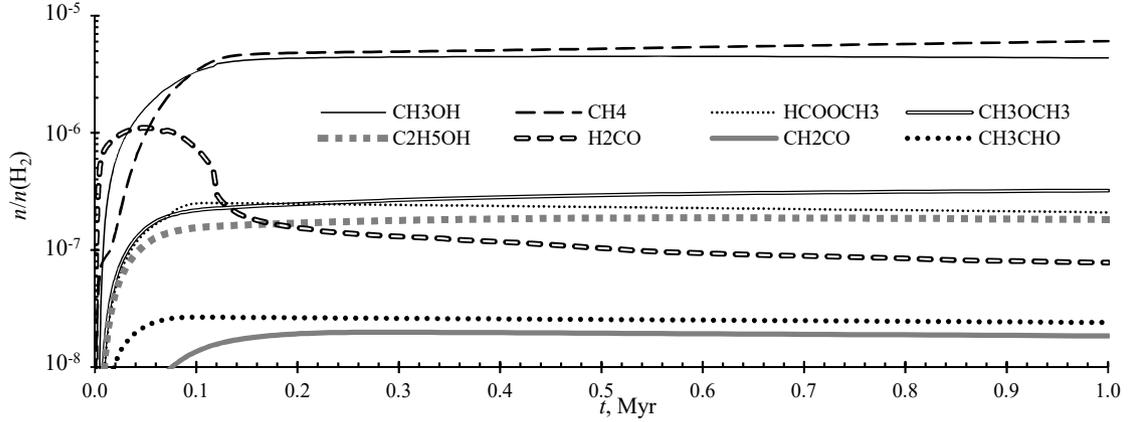}
		\vspace{-20cm}
		\caption{Combined (small + big grain) calculated abundances relative to H$_2$ of frozen COMs and other organic molecules in the icy mantles of interstellar grains in the Standard model.}
    \label{fig-ice}
\end{figure*}

\begin{figure*}
	\includegraphics[width=20cm]{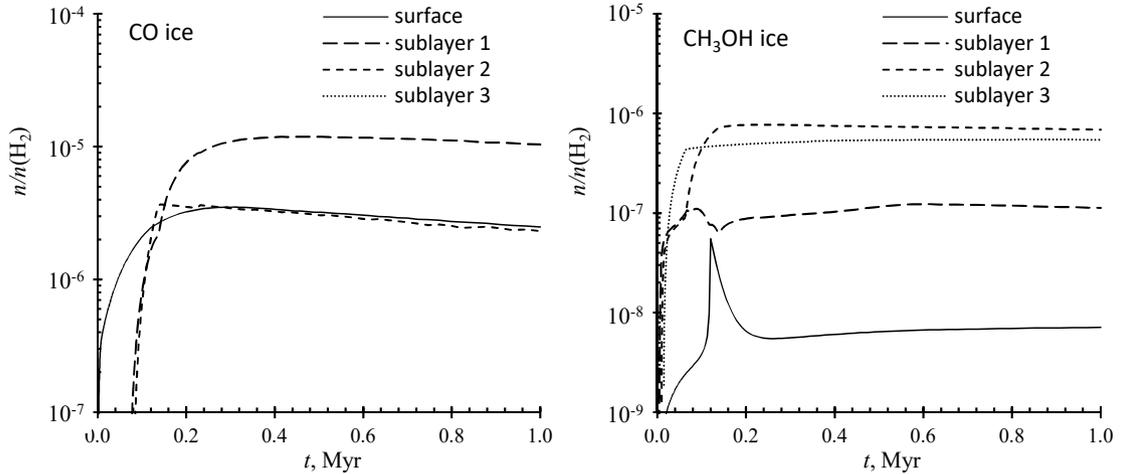}
		\vspace{-20.5cm}
		\caption{Calculated abundances of carbon monoxide and methanol relative to H$_2$ in the four ice layers on the small 0.01\,$\mu$m grains the Standard model. Sublayers 1, 2, and 3 represent increasingly deeper ice layers (closer to grain core). CO is almost nonexistent in sublayer~3 and does not appear in the plot.}
    \label{fig-layer}
\end{figure*}

Our model describes the icy mantles on grains as consisting of four sublayers. While there are models that resolve ice structure in more detail \citep{Chang16,Furuya17}, our approach is sufficient for the aims of this study. In the present model, each sublayer is up to about 8\,MLs thick. As explained in Section~\ref{m_colli}, in the Standard model, collisions remove a spherical cap consisting of the surface and the shallowest sublayer from the small grains. In the Maximum model, an ice volume equal to a spherical cap consisting of all four layers is detached, while the No-collision model does not consider collisional desorption. Ice composition in all three models is similar and is displayed in Figure~\ref{fig-major}.

As discussed in the previous section, the Standard model produces only minor increase in the abundance of gaseous COMs, relative to the No-collision model, while the Maximum model has significantly higher COM abundances (Table~\ref{tab-com}). The explanation for this result is several-fold. First, grain collisions occur five times as often in the Maximum model. Second, Figure~\ref{fig-layer} shows that the deeper sublayers 2 and 3 are richer in methanol than the surface and its adjacent sublayer~1. Third, the volume of the spherical cap (and thus, the total number of molecules) pushed off the grain is about eight times higher in the Maximum model.

Figure~\ref{fig-ice} shows the calculated abundances of COMs and related organic molecules in ices. The dominant icy COM is methanol, which, along with methane serves as a precursor for heavier COMs, such as ethanol. The unsaturated aldehyde compounds, prominent in gas-phase chemistry (see above section), can be easily hydrogenated and are less abundant in ices. Figure~\ref{fig-layer} shows that CO and CH$_3$OH ice abundance in the layers are somewhat counter-related -- methanol is most abundant in the inner sublayers 2 and 3, which have low CO abundances. This is because photodissociation of H$_2$O in these layers results in hydrogenation and oxidation of CO, converting it to CO$_2$, CH$_3$OH and the heavier COMs. The synthesis of COMs in ices in a similar model is discussed in detail by \citet{K15apj2}.

\section{Conclusions}
\label{cncl}

Summarizing, we have employed an astrochemical model with constant physical conditions, two grain sizes, and multilayer approach on ice mantle modelling to explore the potential efficiency of mechanical collisional desorption of ice fragments from grains. The fragments were assumed to quickly disintegrate into their constituent molecules. While we find that grain collisions due to ambipolar diffusion in truly quiescent cores are unable to induce significant desorption, the opposite is true for turbulence-induced collisions with a grain relative velocity of 50\,m\,s$^{-1}$.

Given that many `starless' cores with detected COMs have nearby active star-forming regions \citep[e.g., L1689b,][]{Pattle21} or are themselves engaged in such activity \citep[e.g., L1544 and Barnard 1-b,][]{Crutcher04,Hirano14}, gas motions and resulting high collisional velocities even exceeding 50\,m\,s$^{-1}$ are not unfeasible in such cores. The same can be said about massive star-forming regions. Interestingly, higher collisional desorption rates are unable to raise the abundance of COMs by more than a factor of 2 in our model. This is because of the unselective desorption of all icy molecules, which increases also the abundance of gaseous OH, CH and other species that destroy methanol and other COMs.

A major approximation of the model is the use of only two grain size bins -- big and small grains. Practically this means that most of the icy molecules are on the small grains, from where they can be readily collisionally desorbed. The real grain size distribution in the interstellar medium is smooth and a substantial part of ice likely resides on middle-sized grains. In order to immediately qualitatively estimate the effects of collisional desorption in an environment with multiple grain sizes, we created a test model that considers five grain types with sizes and abundances taken from \citet{Acharyya11}. Multiple grain sizes may affect several aspects of gas and surface chemistry. Here, we are interested in the possibility that middle-sized grains may not have sufficient velocities relative to smaller or larger grains to induce significant collisional destruction of icy mantles. This aspect can be especially true in the case of ambipolar diffusion relevant to prestellar cores \citep{Silsbee20}. Other properties of this five-grain-size model were retained similar to our base model (Section~\ref{m_mtd}). The test showed that the abundance of gaseous methanol increases by a factor of 40 when desorption occurs only from the smallest grains colliding only with the largest grains (Case~1), and by a factor  of 250 when desorption occurs from the two smallest grain sizes (Case~2). For simplicity, here we assumed collision speed always to be 50\,m\,s$^{-1}$. The attained CH$_3$OH gas phase abundances after 1\,Myr are $9\times10^{-12}$ and $5\times10^{-11}$ relative to H$_2$ for Cases 1 and 2, respectively. The abundances of other COMs are lower but still comparable to those in the Maximum model.

For an environment with multi-sized grains, turbulence induces collisions with appreciable velocities even between grains of the same size \citep{Ormel07}. Such collisions means that both, the Maximum main model and the above-described Case~2 supplementary model are conservative estimates of the efficiency of collisional desorption. Summarizing, one can conclude that such desorption may indeed be an important also in a multi-sized grain environment.

To our knowledge, the closest recent work attempting to explain the observations of COMs in cold cores is that of \citet{Vasyunin17}, who investigate reactive (chemical) desorption as the prime culprit. Compared to the reactive desorption, collisional desorption has two features that makes it more feasible. First, thanks to collisions, COMs have elevated abundances up to $\approx1$\,Myr time-scale, which may explain the numerous detections of COMs in starless cores by \citet{Scibelli20}. In other words, collisions are able to sustain high COM abundances for a prolonged time, not just achieve an elevated abundance during a desorption efficiency peak. Second, the high COM abundances were achieved for an ice composition that is typical for starless cores (CH$_3$OH:H$_2$O is 2.7 per cent in the Maximum model). In a case when methanol abundance in the gas is approximately proportional to its abundance in ice, collisional desorption may help to explain the observed abundances of COMs in the L1544 core, where CH$_3$OH:H$_2$O has been observed to be 11 per cent \citep{Scibelli21}. On the other hand, \citet{Vasyunin17} achieve their agreement with observations by simulating a high methanol ice abundance (28 per cent of H$_2$O ice) that has not been observed in starless cores.

The above features indicate that grain collisions can be a candidate mechanism for explaining the observations of COMs in starless cores. The efficiency of collisional desorption can be affected by consideration of a more realistic grain size distribution, along with temperature and cosmic-ray induced desorption rate that is specifically adjusted for each grain size. Such approach leads to variations in ice thickness and composition between grain populations with various sizes \citep{Pauly16,Silsbee21}. Another aspect that can be improved is the description of icy grain collision mechanics (Section~\ref{m_colli}). For example, we have assumed a rather conservative approach on calculating the collisional cross section (Equation~(\ref{colli4})), while it can be reasoned that detachment of non-polar icy fragments occurs also in head-on collisions if the icy mantles are shattered.

Finally, our network is based on UDfA12, with only a few additions. Such a standardized reaction list is good for replicable results but describes the chemistry of COMs only to a limited degree. This limitation may explain some of the disagreements in the comparison of calculated and observed abundances. However, it is encouraging that such generally close results were achieved with a standard network. It is a testament to the credibility of UDfA12 and to the potential of the mechanical collisional desorption mechanism for applications with a more detailed network.

Assuming that collisional desorption is active, there are a few indirect conclusions from this study:
\begin{itemize}
	\item collisional desorption is more efficient from small icy grains, which may fuel accumulation of ices on the big grains in the long-term;
	\item the removal of icy mantle fragments may lead to small icy grains having irregular shapes;
	\item the sticking of interstellar grains covered by soft and volatile ice layers may occur at higher speeds because a part of the collision energy can be consumed by desorbing molecules and ejecting icy fragments;
	\item abundant gaseous COMs in a cold core may be a signature for turbulence, influx of matter, or other gas motions that induce dust collisions in such cores.
\end{itemize}

\section*{Acknowledgements}
JK has been funded by Latvian Science Council project No. lzp-2018/1-0170 “Evolution of Organic Matter in the Regions of Star and Planet Formation (OMG)”. JK also is grateful to Ventspils Town Council for support.
\section*{Data Availability}
The data underlying this article will be shared on reasonable request to the corresponding author.



\bibliographystyle{mnras}
\bibliography{Collide} 

\bsp	
\label{lastpage}
\end{document}